\documentclass[pre,twocolumn,showpacs,square,amssymb,amsmath,nobibnotes]{revtex4-1}
\usepackage{bm}
\usepackage{times}
\usepackage{graphicx}
\usepackage[usenames,dvipsnames,svgnames,table]{xcolor}
\usepackage{hyperref}
\usepackage{physics}
\hypersetup{colorlinks=true,  linkcolor=NavyBlue, citecolor=PineGreen,urlcolor=cyan}
\usepackage{physics}
\usepackage{comment}
\usepackage{xcolor}

\hypersetup{colorlinks=true, linkcolor=NavyBlue, citecolor=PineGreen,urlcolor=cyan}

\newcommand{\erfc}{{\rm erfc}}

\newcommand{\bracket}[1]{\left\langle #1\right\rangle}
\newcommand{\beeq}[1] {\begin{equation}\begin{split}#1\end{split}\end{equation}}

\newcommand{\nn}{\nonumber}
\newcommand{\be}{\begin{equation}}
\newcommand{\ee}{\end{equation}}
\DeclareMathOperator{\sign}{sign}

\begin{document}

\title{Intermediate deviation regime for the full eigenvalue statistics  in the complex Ginibre ensemble}

\author{Bertrand Lacroix-A-Chez-Toine}
\affiliation{LPTMS, CNRS, Univ. Paris-Sud, Universit\'e Paris-Saclay, 91405 Orsay, France}

\author{Jeyson Andr\'es  Monroy Garz\'on}
\author{Christopher Sebastian Hidalgo Calva}
\affiliation{Departamento de F\'isica Cu\'antica y Fot\'onica, Instituto de F\'isica, UNAM, P.O. Box 20-364, 01000 Mexico Distrito Federal, Mexico}

\author{Isaac P\'erez Castillo}
\affiliation{Departamento de F\'isica Cu\'antica y Fot\'onica, Instituto de F\'isica, UNAM, P.O. Box 20-364, 01000 Mexico Distrito Federal, Mexico}
\affiliation{London Mathematical Laboratory, 18 Margravine Gardens. London W6 8RH, United Kingdom}

\author{Anupam  Kundu}
\affiliation{International Centre for Theoretical Sciences, TIFR, Bangalore 560089, India}

\author{Satya N. Majumdar}
\author{Gr\'egory Schehr}
\affiliation{LPTMS, CNRS, Univ. Paris-Sud, Universit\'e Paris-Saclay, 91405 Orsay, France }

\begin{abstract}
We study the Ginibre ensemble of $N \times N$  complex  random matrices and compute exactly, for any finite $N$, the full distribution as well as all the cumulants of the number $N_r$ of eigenvalues within a disk of radius $r$ centered at the origin. In the limit of large $N$, when the average density of eigenvalues becomes uniform over the unit disk, we show that for $0<r<1$ the fluctuations of $N_r$  around its mean value $\langle N_r \rangle \approx N r^2$ display three different regimes: (i) a typical Gaussian regime where the fluctuations are of order ${\cal O}(N^{1/4})$, (ii) an intermediate regime where $N_r - \langle N_r \rangle = {\cal O}(\sqrt{N})$, and (iii) a large deviation regime where $N_r - \langle N_r \rangle = {\cal O}({N})$. This intermediate behaviour (ii) had been overlooked in previous studies and we show here that it ensures a smooth matching between the typical  and the large deviation regimes. In addition, we demonstrate that this intermediate regime controls all the (centred) cumulants of $N_r$, which are all of order ${\cal O}(\sqrt{N})$, and we compute them explicitly. Our analytical results are corroborated by precise ``importance sampling'' Monte Carlo simulations.  
\end{abstract}
\maketitle

%%%%%%%%%%%%%%%%%%%%%%%%%%
\section{introduction}
%%%%%%%%%%%%%%%%%%%%%%%%%%
Since the seminal works of Wishart \cite{wishart1928generalised}, in statistics, and Wigner \cite{wigner1958distribution}, in nuclear physics,  there has been an ever growing number of topics where random matrix theory has found useful applications \cite{mehta2004random,forrester2010log,akemann2011oxford}. This encompasses quantum chaos~\cite{bourgade2013quantum}, mesoscopic transport \cite{beenakker1997random}, stochastic growth models~\cite{majumdar2007courseH,kriecherbauer2010pedestrian}, trapped fermions \cite{marino2014phase,castillo2014spectral,dean2019noninteracting} or disordered systems \cite{biroli2007extreme} in physics, as well as combinatorics or number theory \cite{keatingleshouches2017} in mathematics, to name just a few. Of crucial importance are the various statistical properties of these random matrices \cite{mehta2004random,forrester2010log}, which for many of these  applications, correspond to deriving  the {\it full counting statistics} (FCS), i.e. the statistics of the number $N_{\cal D}$ of eigenvalues of a random matrix 
contained in a given domain~${\cal D}$. This domain might be an interval on the real line for rotationally invariant ensembles, like the $\beta$-Gaussian ensemble for instance, or a region of the complex plane for non-invariant ones, like the Ginibre ensemble. {Questions related to FCS in random matrix ensembles and related Coulomb gas models have actually attracted a lot attention both in physics (see e.g. \cite{torquato2003local,majumdar2009index,majumdar2011many,hexner2015hyperuniformity}) and mathematics (see e.g. \cite{chatterjee2017rigidity,leble2018fluctuations}) during the last few years. In particular, it was realised that these random systems exhibit some ``rigidity'' or ``hyperuniformity'' (for reviews see \cite{ghosh2017fluctuations, torquato2018hyperuniform}): in these cases, indeed, the variance of $N_{\cal D}$ usually grows slower than the volume of the domain~${\cal D}$.}

The most studied case concerns $N \times N$ random matrices belonging to the Gaussian Unitary Ensemble (GUE). In this case, the average density of eigenvalues $\rho_N(\lambda)$ approaches the celebrated Wigner's semi-circle law~$\rho_N(\lambda)~\approx~\sqrt{2 - \lambda^2}/\pi$~\cite{wigner1958distribution}, which has a compact support $[-\sqrt{2}, +\sqrt{2}]$. For a domain located deep inside the 
bulk of the spectrum, i.e. for a segment ${\cal D} = [-L,L]$ around the origin and of size comparable to the inter-particle distance, 
it is well known from Dyson's and Mehta's works \cite{dyson1962statistical1,dyson1962statistical2,dyson1962statistical3,dysonmehta} that the variance of $N_{\cal D}$ scales logarithmically with the size of ${\cal D}$. Later on, it was shown that the typical fluctuations of $N_{\cal D}$, in the bulk, are Gaussian \cite{costin1995gaussian,fogler1995probability,soshnikov2000gaussian}, while the large deviations in the bulk regime were studied in \cite{holcombvalko}. It is only recently \cite{marino2014phase,marino2016number} that the fluctuations of $N_{\cal D}$ for an interval of size larger than the bulk, i.e. for $L = {\cal O}(1)$, were investigated, motivated to a large extent by the connection between the GUE and the problem of $N$ non-interacting fermions in a harmonic trap at zero temperature. Interestingly, it was shown that the variance exhibits a non-monotonic behavior: after a logarithmic growth $\propto \ln(N\,L)$ in the ``extended bulk'', i.e. for $0 < L < \sqrt{2}$, it decreases abruptly as $L$ approaches the edge $L \approx \sqrt{2}$, finally decaying exponentially fast (with $N$) for $L > \sqrt{2}$. Furthermore, the full probability distribution of $N_{\cal D}$, $P_{\cal D}(K,N) = {\rm Prob}(N_{\cal D} = K)$, 
was computed for large $N$ using Coulomb gas techniques \cite{marino2014phase,marino2016number}, and it was found that, in the extended bulk regime, it follows a large deviation principle of the form 
\be \label{large_dev}
P_{\cal D}(K,N) \approx e^{- N^2 \psi_L(K/N)} \;,
\ee 
where the function $\psi_L(x)$ was obtained explicitly. In standard large deviation forms, the rate function $\psi_L(x)$ usually has a quadratic behavior around its minimum $x_{\min}$, indicating that the typical fluctuations are described by a Gaussian distribution. In this case, however, it was first shown for the distribution of the index number of the GUE \cite{majumdar2009index,majumdar2011many}, that this quadratic form is modulated by a logarithmic singularity, i.e. $\psi_L(x_{\min} + \delta) \propto \delta^2/|\ln \delta|$ as $\delta \to 0$. A similar behavior was later found for the index distribution of Cauchy random matrices \cite{marino2014index}. These results  indicate that there are actually two scales associated to the fluctuations of $N_{\cal D}$: (i) a short scale, $N_{\cal D} - \langle N_{\cal D}\rangle = {\cal O}(\sqrt{\ln {N L}})$, which describes the typical fluctuations that are Gaussian and (ii) a larger scale, $N_{\cal D} - \langle N_{\cal D}\rangle = {\cal O}(N)$ associated to atypical fluctuations, which are no longer Gaussian but instead described by the large deviation principle given by Eq. \eqref{large_dev}. The matching between these two regimes is ensured by the logarithmic singularity of the rate function $\psi_I(x)$ near its minimum. A similar scenario was later shown to hold for all the classical rotationally invariant ensembles of random matrix theory (RMT), including the $\beta$-Gaussian, the $\beta$-Wishart and the $\beta$-Cauchy ensembles \cite{marino2016number}. 

What about the case of non-invariant ensembles of random matrices? To answer this question, we focus on the complex Ginibre ensemble \cite{mehta2004random,forrester2010log,ginibre1965statistical}, which corresponds to the set of $N\times N$ random matrices $\bm{M}$ with independent and identically distributed (i.i.d.) complex normal random entries with real and imaginary parts of variance $1/(2N)$. The $N$ eigenvalues are complex and their joint probability distribution function (PDF) is given by
\beeq{
P_{\text{joint}}(z_1,\ldots,z_N)=\frac{1}{Z_N}\prod_{i<j}|z_i-z_j|^2\prod_{k=1}^N e^{-NV(z_k)}\label{P_joint}\,,
}
with $V(z)=|z|^2$, and $Z_N$ being a normalization factor. This ensemble  appears in a variety of contexts, ranging from chaotic dissipative quantum systems \cite{haake2013quantum}, the two dimensional one-component plasma \cite{forrester1998exact,cunden2016large}, the  study of normal random matrices (i.e. matrices ${\bm M}$ that satisfy the commutation relation $[\bm{M},\bm{M}^\dag]=0$ \cite{chau1992unitary,chau1998structure,ameur2011fluctuations,lacroix2018extremes}). More recently, it was shown that the eigenvalues $z_i$ of a complex Ginibre matrix are in one-to-one correspondence with the positions of non-interacting fermions in a two dimensional {\it rotating} harmonic potential \cite{lacroix2019rotating}. In this context, the study of the FCS is of great interest, since in many non-interacting fermionic systems the FCS in a given domain ${\cal D}$ is often related to the entanglement entropy of ${\cal D}$ with the rest of the system \cite{klich2006lower,song2011entanglement,calabrese2015random}, which for this system was recently studied in \cite{lacroix2019rotating} using this connection to RMT.

\begin{figure}[t!]
\begin{center}
\includegraphics[width = 0.7\linewidth]{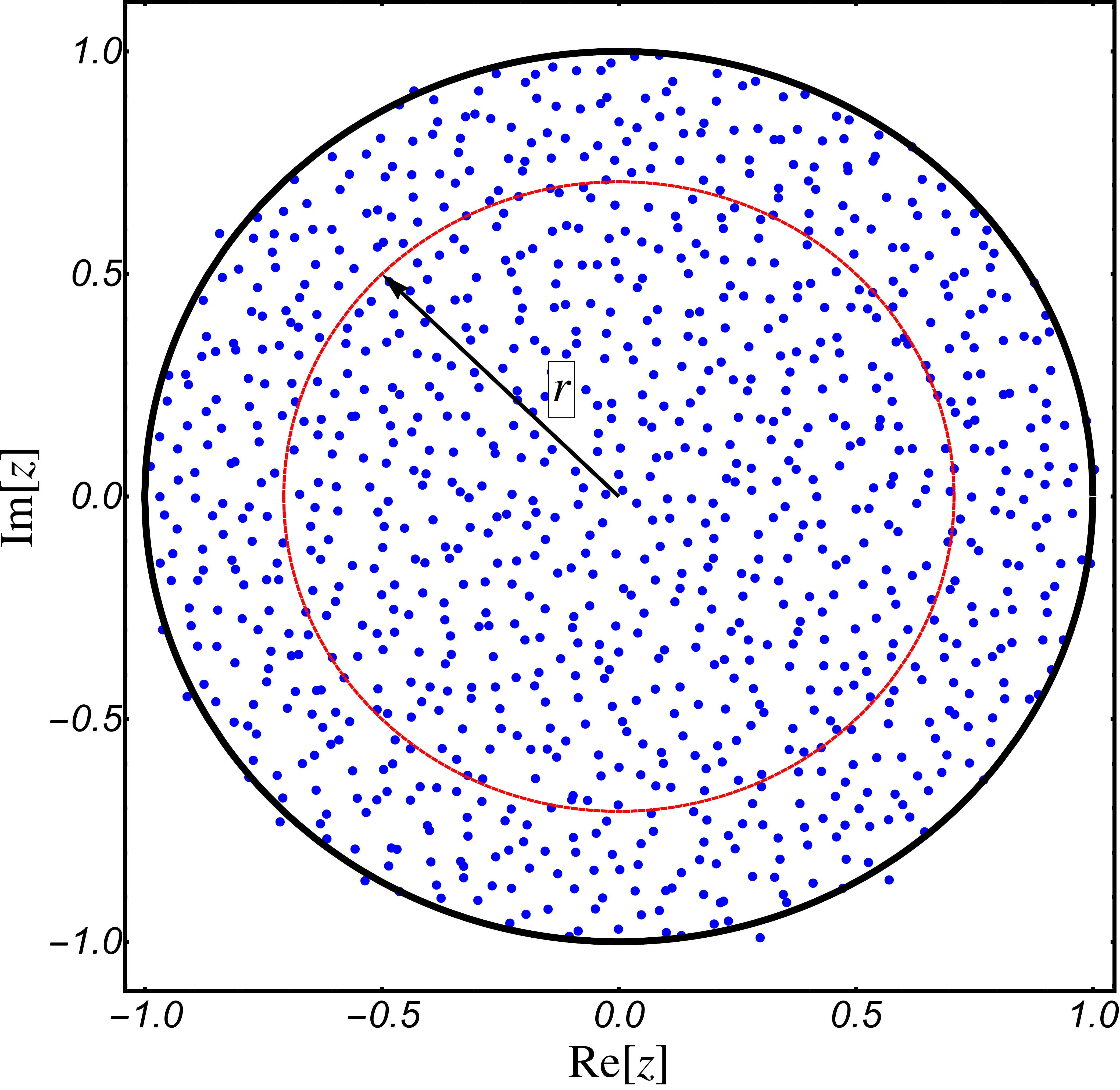}
\caption{Snapshoot of the position of eigenvalues of the Ginibre ensemble generated by Monte Carlo simulations for $N=100$.}
\label{fig:0}
\end{center}
\end{figure}

For the  Ginibre ensemble given by Eq. (\ref{P_joint}) with $V(z) = |z|^2$, in the limit of large $N$, the limiting density is uniform on the unit disk, i.e. $\rho_N(z) \approx (1/\pi) \Theta(1 -|z| )$, where $\Theta(x)$ is the Heaviside step function. Clearly, the typical distance between two neighbouring eigenvalues is ${\cal O}(1/\sqrt{N})$. In the following, we focus on the FCS inside a disk of radius $r$ centered at the origin, i.e. ${\cal D} = \{z \in \mathbb{C}, |z| \leq r \}$ with $0<r<1$. We denote  the number of eigenvalues inside this disk by $N_r$ (see Fig. \ref{fig:0}). Our main goal is to compute the probability $P_r(K,N) = {\rm Prob}(N_r= K)$. 
Since we naturally expect that $N_r \propto N$, we find convenient to work with the rescaled variable $\kappa=N_r/N$ and introduce its corresponding  probability distribution 
\be
P_r(K,N) = \frac{1}{N} {\cal P}_{r}\left(\kappa = \frac{K}{N} , N\right) \label{def_scaled_proba} \;.
\ee
In Ref. \cite{allez2014index}, the function ${\cal P}_r(\kappa,N)$ was studied in the limit of large $N$ with $0<r<1$ and it was shown to take a scaling form similar to the one given by Eq. \eqref{large_dev}, that is
\be \label{large_dev_Ginibre}
{\cal P}_r(\kappa,N) \approx e^{-N^2 \Psi_r(\kappa)}\,,
\ee
where the rate function $\Psi_r(\kappa)$ was computed explicitly using Coulomb gas techniques \cite{allez2014index}, yielding
\beeq{\label{explicit_psi}
\Psi_r(\kappa)=\frac{1}{4}\left|(r^2-\kappa)(r^2-3 \kappa)-2\kappa^2\log\left(\frac{\kappa}{r^2}\right)\right|\,,
}
with $0 \leq \kappa \leq 1$. Note that a very similar large deviation form (\ref{large_dev_Ginibre}), with the same rate function (up to trivial rescalings) was also obtained in the bulk, i.e. for $r = {\cal O}(1/\sqrt{N})$ in Ref.~\cite{shirai2006large}. Since the density is uniform, one immediately has  that $\langle N_r \rangle = N\, r^2$. This can also be checked  from Eq. (\ref{large_dev_Ginibre}), since  $\Psi_r(\kappa)$ has a minimum at $\kappa_{\min} = r^2$. However,  contrary to  the GUE case, the rate function has a {\it cubic} behavior near the minimum, i.e.
\be \label{cubic_behavior}
\Psi_r(\kappa) \approx \frac{|\kappa-r^2|^3}{6 r^2} \;, \; \kappa \to r^2 \;.
\ee  
This cubic behavior is rather surprising as one would na\"ively expect the typical fluctuations of $N_r$  to be  Gaussian, as it  was rigorously shown in the bulk when $r= {\cal O}(1/\sqrt{N})$ \cite{soshnikov2000determinantal, shirai2006large}. Oddly enough, the latter would imply a {\it quadratic} behavior of the rate function, instead of the cubic one  given by  Eq. \eqref{cubic_behavior}.  Besides, this cubic behavior would further suggest that the typical fluctuations of $N_r$ are of order ${\cal O}(N^{1/3})$ \cite{allez2014index}, which  contradicts  a recent computation \cite{lacroix2019rotating} where it was shown  that the typical fluctuations of $N_r$ are instead of order ${\cal O}(N^{1/4})$. In short: the behavior encapsulated in Eq. \eqref{cubic_behavior} is thus quite puzzling.

In this paper, we solve this puzzle and show that, at variance with the GUE case, there exists in the present case {\it three} different scales associated to the fluctuations of $N_r - \langle N_r\rangle$: (i) a shorter scale $\mathcal{O}(N^{1/4})$ corresponding to the typical fluctuations which are indeed Gaussian (as in the bulk \cite{shirai2006large}), (ii) an {\it intermediate} scale of $\mathcal{O}(\sqrt{N})$ associated to ``moderate'' deviations which are non-Gaussian (see below) and finally (iii) a larger scale $\mathcal{O}(N)$ corresponding to large atypical fluctuations, which are described by the large deviation form given by Eqs. \eqref{large_dev_Ginibre} and \eqref{explicit_psi}. Our findings can be summarised as follows 
\beeq{ \label{main_results}
{\cal P}_r(\kappa,N)\approx
	\begin{cases}
	\displaystyle		e^{-N^{\frac{3}{2}}\frac{\sqrt{\pi}}{2r}(\kappa-r^2)^2}&\;,\;|\kappa-r^2|\sim N^{-\frac{3}{4}}\\
	\displaystyle		e^{-\sqrt{2N}r\Psi_{I}\left(\sqrt{\frac{N}{2r^2}}(\kappa-r^2)\right)}&\;,\;|\kappa-r^2|\sim N^{-\frac{1}{2}}\\
	\displaystyle	e^{-N^2 \Psi_r(\kappa)}&\;,\;|\kappa-r^2| \sim N^0
	\end{cases}
}
where we remind that $\kappa=N_r/N$. Note that the fluctuations of $N_r$ is $N$ times larger than the fluctuations of $\kappa$.
%Note that, as we may sometimes discuss results in terms of $\kappa$ rather than $N_r$, the fluctuations of the latter correspond to $N$ times the ones of the former. 
Our main result is thus the existence of an intermediate regime of fluctuations [the second line in (\ref{main_results})], which had been overlooked in previous studies \cite{shirai2006large,allez2014index}. This regime is characterized by the 
rate function $\Psi_I(x)$ given in Eqs. (\ref{chi_mu_inter2}) and (\ref{PsiI}) below. In the asymptotic regimes of the argument, this rate function behaves as
\be\label{psi_asympt}
\Psi_{I}(x)\approx \begin{cases}
\displaystyle\sqrt{\frac{\pi}{2}}x^2&\;,\;\;x\to 0\;,\\
&\\
\displaystyle\frac{1}{3}\,|x|^3&\;,\;\;x\to \pm \infty
\end{cases}\,.
\ee
Hence the quadratic form in the first line of (\ref{psi_asympt}) matches smoothly with the central Gaussian behavior while the cubic expression in the second line of Eq. (\ref{psi_asympt}) matches with the cubic behavior of the large deviation function (\ref{cubic_behavior}) near its minimum (see Appendix \ref{AppE} for details). Moreover, this intermediate deviation regime turns out to be extremely important since, as we later show, it controls all the cumulants of $N_r$ beyond the first one (at the leading order for large $N$). These read  
\be\label{cumulants}
\langle N_r^p\rangle_c \approx - \sqrt{2N}\,r \int_{-\infty}^\infty dx \, {\rm Li}_{1-p}\left(- \frac{\erfc(x)}{\erfc(-x)} \right) \;,
\ee
with $p>1$ and where ${\rm Li}_n(x) = \sum_{k \geq 1} x^k/k^n$ is the polylogarithm function of index $n$. Using the identity ${\rm Li}_{-n}(x) = (-1)^{n+1} {\rm Li}_{-n}(1/x)$, for $n$ a positive integer, it is easy to see that the odd order cumulants in Eq. (\ref{cumulants}) vanish at the leading  order ${\cal O}(\sqrt{N})$. In fact we argue (see Eqs.~\eqref{general_Lk}-\eqref{chi_mu_univ}) that, for the Coulomb gas model in Eq. (\ref{P_joint}) with arbitrary spherically symmetric potential $V(z) = v(|z|)$ (such that $v(r) \gg \ln r^2$ for $r \gg 1$), the cumulants of $N_r$ also behave, for large $N$, as those given by Eq. (\ref{cumulants}), up to a $v$-dependent prefactor [see Eq. (\ref{chi_mu_univ}) below]. Finally, we point out that the very same rate function $\Psi_I(x)$ shows up in a completely different problem, that of  the large deviation function of the particle current through the origin in one-dimensional diffusing particles \cite{derrida2009current}.

This paper is organised as follows: in Section \ref{sect:definitions} we introduce the complex Ginibre ensemble and derive an exact expression for the cumulant generating function of $N_r$, valid for any finite $N$. In Section \ref{sect:results} we analyse this exact result in the asymptotic limit of large $N$ and describe in detail the three different regimes of typical, intermediate and atypical fluctuations. In Section \ref{sect:DerGer} we discuss the occurrence of the intermediate deviation function $\Psi_I(x)$ in diffusive systems, and, in Section \ref{sect:MC} we compare our analytical results to (reweighted) Monte Carlo simulations, before summarising our findings and pointing out some remarks  in Section \ref{sect:conclusions}. Some details of the derivations have been relegated to the appendices.

%%%%%%%%%%%%%%%%%%%%%%%%%%
\section{Exact results for finite $N$}
\label{sect:definitions}
%%%%%%%%%%%%%%%%%%%%%%%%%%
We start with the  joint PDF in Eq. (\ref{P_joint}) with a generic spherically symmetric potential $V(z)=v(|z|)$. Let $r_i=|z_i|$ be the modulus of the eigenvalue $z_i$. The joint distribution of the $r_i$'s can be computed explicitly from Eq. \eqref{P_joint} by integrating over the phases $\theta_i=\arg{z_i}$ of the eigenvalues. This yields~\cite{kostlan1992spectra,chafai2014note} (see also Appendix \ref{app_CGF}) 
\begin{eqnarray}
P_{\rm rad}(r_1,\cdots,r_N)&=&\int_0^{2\pi}\cdots\int_0^{2\pi} \prod_{i=1}^N r_i \, d\theta_i \, P_{\text{joint}}(z_1,\ldots,z_N)\nn\\
&=&\frac{1}{N!}\sum_{\sigma\in {\cal S}_N}\prod_{k=1}^N \frac{r_k^{2\sigma(k)-1}}{h_{\sigma(k)}}e^{-Nv(r_k)}\;,\label{P_rad}
\end{eqnarray}
where ${\cal S}_N$ denotes the permutation group of $N$ elements while the coefficients $h_k$'s are given by
\be
h_k=\int_0^{\infty} r^{2k-1}e^{-N v(r)}dr\;. \label{h_k}
\ee
The expression of $P_{\rm rad}(r_1,\cdots,r_N)$  is manifestly invariant under any permutation of the $r_i$'s (note that it can also be written as a permanent). 

We are interested to compute the number statistics inside a disk of radius $r$ and centered at the origin, i.e. the number of eigenvalues $N_r$ inside this disk 
\beeq{
N_r = \sum_{i=1}^N\Theta(r-r_i)\,\label{index} \;.
}
Starting from the expression  \eqref{P_rad}, one can easily compute the moments of $N_r$. For instance, 
using the invariance property of the joint distribution in Eq.~\eqref{P_rad} under permutation of $r_i$'s, one can easily show that 
the mean is given by 
\begin{eqnarray}\label{av_Nr}
\hspace*{-0.5cm}\langle N_r \rangle =  \sum_{k=1}^N L_{k}(r)\;, \; L_{k}(r)=\frac{1}{h_k}\,\int_0^r du \, u^{2k-1}e^{-Nv(u)} ,
\end{eqnarray}
where the notation $\bracket{(\cdots)}$ corresponds to  averaging with respect to the joint PDF given by Eq. \eqref{P_rad}. To derive both the full distribution of $N_r$ and a compact expression for higher order cumulants, it is more convenient to  introduce the (centered) cumulant generating function (CGF) $\chi_r(\mu,N)$ as
\begin{eqnarray}
\chi_r(\mu,N) = \ln\bracket{e^{-\mu (N_r- \langle N_r \rangle)}} \;.
\end{eqnarray}
Since $N_r$, as defined in Eq. \eqref{index}, is a symmetric function of the moduli $r_i$'s, one can easily obtain the following expression for the CGF
\begin{eqnarray}
\hspace*{-0.5cm}\chi_r(\mu,N)=\sum_{k=1}^N \left(\log \left[ M_{k}(r)  + e^{-\mu} L_{k}(r)\right] + \mu L_{k}(r) \right),
\label{eq: CGF}
\end{eqnarray}
where $M_{k}(r) = 1 - L_{k}(r)$. Furthermore, from $\chi_r(\mu,N)$, one can formally derive an explicit expression for the probability $P_r(K,N) = {\rm Prob}(N_r = K)$ of having  exactly $K$ eigenvalues inside the disk of radius $r$ centered at the origin, with $0 \leq K \leq N$. Indeed, noting that $\langle e^{- \mu N_r } \rangle$ can also be written as the generating function of the variable $z=e^{-\mu }$ of the probabilities $P_r(K,N)$ with respect to $K$, one has that
\be
\langle e^{- \mu N_r }\rangle = \sum_{K=0}^{N} e^{-\mu K}P_r(K,N)\;.\label{GF} 
\ee
From Eq. (\ref{eq: CGF}), we see that $\langle e^{- \mu N_r }\rangle$ is a simple polynomial of degree $N$ in the variable $z = e^{-\mu}$, that is $\langle e^{- \mu N_r }\rangle = \prod_{k=1}^N \left[ M_{k}(r) + e^{-\mu} L_{k}(r)\right]$. Therefore we obtain the explicit expression
\be\label{explicit_PKN}
P_r(K,N) = \left[\prod_{k=1}^N M_{k} \right] e_K\left(\frac{L_{1}}{M_{1}}, \frac{L_{2}}{M_{2}}, \cdots, \frac{L_{N}}{M_{N}} \right)\,,
\ee
where we used the shorthand notations $L_k \equiv L_k(r)$, $M_k \equiv M_k(r)$, and where
\be \label{symetric_poly}
e_K(x_1, \cdots, x_N) = \sum_{1\leq \lambda_1 < \cdots < \lambda_K \leq N} \prod_{k=1}^K x_{\lambda_k}
\ee
is the elementary symmetric polynomial of $N$ variables and degree $K$. In addition, by expanding Eq. (\ref{eq: CGF}) in powers of $\mu$, it is possible to compute all the cumulants $\langle N_r^p\rangle_c$ of order $p>1$, yielding \cite{lacroix2019rotating}
\begin{eqnarray} \label{cumul_exact}
\langle N_r^p\rangle_c = -\sum_{k=1}^N {\rm Li}_{1-p} \left( -\frac{L_k(r)}{M_k(r)}\right)\,,
\end{eqnarray}
where  ${\rm Li}_n(x)$ is the polylogarithm function. 

The formulae (\ref{explicit_PKN}) and (\ref{cumul_exact}) are exact for any $N$ and for any  spherically symmetric potential $v(r)$, the dependence on the latter being contained in the expressions $L_k(r)$ and $M_k(r)  = 1 - L_{k}(r)$ given by Eq. (\ref{av_Nr}). In particular, for the complex Ginibre ensemble, which corresponds to $v(r)=r^2$, we have that
\beeq{\label{Mk_Ginibre}
M_{k}(r)=\frac{\Gamma(k,Nr^2)}{\Gamma(k)}\,,\quad L_{k}(r)=1-\frac{\Gamma(k,Nr^2)}{\Gamma(k)}\,,
}
where $\Gamma(k,z)=\int_z^{\infty}t^{k-1}e^{-t}dt$ is the upper incomplete Gamma function. Note that for $K=N$ and $K=0$, the formula (\ref{explicit_PKN}) becomes fairly simple. In fact, it is easy to see that  $P_r(N,N)={\rm Prob}\left[r_{\max}\leq r\right]$ and $P_r(0,N)={\rm Prob}\left[r_{\min}\geq r\right]$, with $r_{\max} = \max_{1 \leq i \leq N} r_i$ and $r_{\min} = \min_{1 \leq i \leq N} r_i$, that is, they correspond to the the cumulative distributions of $r_{\max}$ and $r_{\min}$, respectively.  Therefore, from Eq.  \eqref{explicit_PKN}, one has that
\begin{align}
P_r(N,N)&={\rm Prob}\left[r_{\max}\leq r\right]=\prod_{k=1}^N L_k(r)\;,\\
P_r(0,N)&={\rm Prob}\left[r_{\min}\geq r\right]=\prod_{k=1}^N M_k(r) \;.
\end{align}
For other values of $K$, the expression of $P_r{(K,N)}$ is a bit more cumbersome but it can be evaluated numerically rather easily. 
We note that in the limit $N \to \infty$ with $r=\mathcal{O}(1/\sqrt{N})$, our formulae (\ref{explicit_PKN}) and \eqref{symetric_poly} yield back the result obtained in Ref. \cite{forrester1992}. 

To analyse the large $N$ behavior of $P_{r}(K,N)$, it turns out that this expression (\ref{explicit_PKN}) is actually of little use. Instead, it is much more convenient to study the large $N$ behaviour of the CGF $\chi_r(\mu,N)$ in Eq. (\ref{eq: CGF}). This  is the purpose of the next section.

%%%%%%%%%%%%%%%%%%%%%%%%%%
\section{The limit of large $N$}
\label{sect:results}
%%%%%%%%%%%%%%%%%%%%%%%%%%

From now on, unless stated otherwise, we focus on the case of Ginibre matrices, corresponding to $v(r)=r^2$ in Eq. (\ref{P_rad}). The starting point of our analysis is the exact expression for the CGF given by Eq. (\ref{eq: CGF}) where $L_k(r)$ and $M_k(r)$ are given in Eq. (\ref{Mk_Ginibre}). We consider separately the three regimes corresponding to (i) typical, (ii) intermediate and (iii) large fluctuations of  $N_r$ away from its mean value~$\langle N_r\rangle$.

%%%%%%%%%%%%%%%%%%%%%%%%%%
\subsection{Regime of typical fluctuations}
%%%%%%%%%%%%%%%%%%%%%%%%%%

In the limit of large $N$, the density of eigenvalue for the complex Ginibre ensemble is uniform over the unit disk, $\rho(r) \approx \frac{1}{\pi}$ and therefore 
\be \label{avN_large}
\langle N_r \rangle \approx N \, r^2\,,
\ee
which can also be obtained from the exact result given in Eq.~(\ref{av_Nr}). To compute the variance, one can extract the coefficient of $\mu^2$ in the small $\mu$ expansion of the CGF $\chi_\mu(r)$ and one obtains the exact formula
\beeq{
{\rm Var}(N_r) = \langle N_r^2\rangle -  \langle N_r\rangle^2 = \sum_{k=1}^N L_k(r)\left(1-L_k(r)\right)\,.
\label{eq:var:exact}
}
In the large $N$ limit, one can show \cite{lacroix2019rotating} that it behaves as 
\be\label{variance_largeN}
{\rm Var}(N_r) \approx \sqrt{\frac{N}{\pi}}r \;.
\ee
{The fact that the variance grows like $\propto r$, i.e. much slower than the area of the disk ($\propto r^2$) demonstrates the hyperuniformity of 
the fluctuations in this system \cite{ghosh2017fluctuations, torquato2018hyperuniform}}. Besides, a more detailed analysis of the CGF allows to show that all the cumulants, other than the first one, grow at most
like $\sqrt{N}$ \cite{lacroix2019rotating}. This indicates that for large $N$ the rescaled variable ${(N_r - \langle N_r \rangle)}/{\sqrt{{\rm Var}(N_r)}}$ converges to a centered Gaussian random variable with unit variance (since its cumulants of order greater than $2$ vanish when $N \to \infty$). Therefore, using  Eqs. (\ref{avN_large}) and (\ref{variance_largeN}), we obtain the result given in the first line of Eq. (\ref{main_results}). Below, in Section \ref{sect:MC}, we compare analytical expressions of the mean (\ref{avN_large}) and variance (\ref{variance_largeN}) with Monte Carlo simulations.

%%%%%%%%%%%%%%%%%%%%%%%%%%
\subsection{Regime of intermediate fluctuations}
%%%%%%%%%%%%%%%%%%%%%%%%%%
The intermediate regime corresponds to taking the limit of large $N$ but keeping $\mu$ fixed in the expression for the CGF $\chi_r(\mu,N)$ given in (\ref{eq: CGF}). In this limit, the discrete sum over $k$ is dominated by the region $k \sim N\,r^2$ and of width $|k - N\,r^2| = {\cal O}(\sqrt{N})$. In this region, $L_k(r)$ can be approximated by the form  \cite{lacroix2018extremes}
\be \label{asympt_Lk_inter}
L_k(r) \approx\frac{1}{2} \erfc\left(\frac{k - Nr^2}{\sqrt{2N\,r^2}} \right) \;.
\ee 
By substituting the asymptotic form (\ref{asympt_Lk_inter}) into Eq. (\ref{eq: CGF}) and replacing the discrete sum over $k$ by an integral over $x = (k-N r^2)/\sqrt{2 N r^2}$, one obtains, for $N \to \infty$ while keeping $\mu$ fixed, that
\be \label{chi_scaling_form_inter}
\chi_r(\mu,N)\approx \sqrt{2N}r\;\chi\left(\mu\right)\,,
\ee
where the scaling function $\chi(\mu)$ is given by
\be\label{chi_mu_inter1}
\chi(\mu) = \int_{-\infty}^\infty dx \, \left( \ln\left [1 + \frac{e^{-\mu}-1}{2} \erfc(x)  \right] + \frac{\mu}{2} \erfc(x)\right) \,.
\ee
It can be easily verified that the integral over $x$ is well defined. Furthermore, by writing the integral over $x$ as $\int_{-\infty}^\infty dx  = \int_{-\infty}^0 dx + \int_0^\infty dx$, performing the change of variable $x \to - x$ in the first integral and using the identity $\erfc(x) + \erfc(-x) = 2$, the scaling function $\chi(\mu)$ can be conveniently written as
\begin{eqnarray}\label{chi_mu_inter2}
\hspace*{-0.5cm}\chi\left(\mu\right)=\int_0^\infty ds \log\left[1+\sinh^2\left(\frac{\mu}{2}\right)\erfc\left(s\right)\erfc\left(-s\right)\right].
\end{eqnarray}
Under this form (\ref{chi_mu_inter2}), we see that $\chi(\mu)$ is an even function of $\mu$, which implies that the
odd cumulants vanish at leading order in this limit. By expanding $\chi(\mu)$ in Eq. (\ref{chi_mu_inter2}) in powers of $\mu$ one can show rather straightforwardly \cite{lacroix2019rotating} that the cumulants of even order are given by the formula (\ref{cumulants}).  For later use, it is also useful to extract the asymptotic behaviors of $\chi(\mu)$.  These read (see Appendix \ref{AppE})
\be\label{asympt_chi}
\chi(\mu)\approx \begin{cases}
\displaystyle\frac{\mu^2}{2\sqrt{2\pi}}&\;,\;\;\mu\to 0\;,\\
&\\
\displaystyle\frac{2}{3}|\mu|^{3/2}&\;,\;\;\mu\to \pm \infty\;.
\end{cases}
\ee
 
 To obtain the PDF of $N_r$, we need to compute the inverse Laplace transform 
\be \label{Brom}
{\cal P}_r(\kappa,N)\approx \int_{\cal C}\frac{d\mu}{2i \pi}e^{N\mu(\kappa-r^2)}e^{\sqrt{2N}r \chi(\mu)}\;,
\ee
where ${\cal C}$ is the Bromwich contour in the $\mu$-complex plane. In the regime $(\kappa-r^2)=\mathcal{O}(N^{-1/2})$, which corresponds precisely to the intermediate regime, the argument of both exponentials in (\ref{Brom}) are of the same order ${\cal O}(\sqrt{N})$ and the Bromwich integral can be evaluated using a saddle-point approximation. This yields the scaling form in the second line of Eq. \eqref{main_results} where the rate function $\Psi_I(x)$ characterising the fluctuations in the intermediate regime is given by the  Legendre transform
\beeq{\label{PsiI}
\Psi_I(x)=-\min_{\mu \in \mathbb{R}}\left\{x\mu+\chi(\mu)\right\}\,, \; x = \sqrt{\frac{N}{2r^2}}(\kappa-r^2) \;.
}
Since the expression \eqref{chi_mu_inter2} of $\chi(\mu)$ is rather involved, it seems fairly cumbersome  to obtain a more explicit expression of the rate function $\Psi_I(x)$. Luckily, $\Psi_I(x)$ can  easily be evaluated numerically from the expression (\ref{PsiI}) (see Section \ref{sect:MC} below). In addition, using the asymptotic behaviors of $\chi(\mu)$ in Eq.~(\ref{asympt_chi}), one can show that the intermediate deviation rate function behaves  asymptotically as in Eq. (\ref{psi_asympt}).

While the previous analysis concerns the Ginibre ensemble with $V(z) = |z|^2$, one can, in fact, show that a similar result can be obtained for a spherically symmetric potential $V(z) = v(|z|)$ so that $v(r)\gg \ln r^2$ as $r\to \infty$. Indeed, for such a potential, the large $N$ behavior of $L_k(r)$ can be obtained by a saddle-point method \cite{lacroix2018extremes}
\be\label{general_Lk}
L_{k}(r)\approx \frac{1}{2}\erfc\left(\sqrt{2\pi N \rho(r)}(u_{k}-r)\right)\;, 
\ee 
where $u_{k}$ is the solution of $u_k v'(u_k)=2k/N$ and $\rho(r)=\frac{1}{4 \pi r}\partial_r\left(r v'(r)\right)$ is the average density of the Coulomb gas. For the quadratic potential $v(r) = r^2$ one has $u_k = \sqrt{k/N}$ together with $\rho(r) = 1/\pi$ and, for $r$ close to $u_k = \sqrt{k/N}$ the formula (\ref{general_Lk}) yields back the one given by (\ref{asympt_Lk_inter}). Inserting the asymptotic form (\ref{general_Lk}) in Eq. (\ref{eq: CGF}) and performing the same manipulations as done for the Ginibre case, one finds that the CGF reads, for large $N$,  
\begin{eqnarray}\label{chi_mu_univ}
\hspace*{-0.5cm}\chi_r(\mu,N) = \ln\bracket{e^{-\mu(N_r-\bracket{N_r})}}\approx\sqrt{2\pi N\rho(r)}\, r\;\chi(\mu)\;,
\end{eqnarray}
with $\chi(\mu)$ being the same function $\chi(\mu)$ as in Eq. (\ref{chi_mu_inter2}). This shows that, in the intermediate regime, the CGF is, for large $N$, universal  up to a multiplicative prefactor $\propto \sqrt{\rho(r)}$ which contains the whole $v$-dependence. An immediate consequence of this  result (\ref{chi_mu_univ}) is that the cumulants of $N_r$ are also universal (up to the same multiplicative pre-factor), i.e. they are given by Eq. (\ref{cumulants}) after  substituting $\sqrt{2N}$ by $\sqrt{2\pi N \rho(r)}$.

%%%%%%%%%%%%%%%%%%%%%%%%%%
\subsection{Regime of atypical fluctuations}
%%%%%%%%%%%%%%%%%%%%%%%%%%

The regime of  atypical fluctuations of $N_r$, for $N_r - \langle N_r \rangle = {\cal O}(N)$ was studied in Ref. \cite{allez2014index} using
a Coulomb gas method, leading to the result in Eqs. (\ref{large_dev_Ginibre}) and (\ref{explicit_psi}). Here we show that this large deviation regime can also be obtained directly from the exact expression for the CGF in  Eq. \eqref{eq: CGF} in the limit of large $N$ with $\mu = {\cal O}(N)$. To do so, we first set $\mu = \lambda N$, with $\lambda = {\cal O}(1)$ for large $N$. Moreover, one can show that, in this limit, the sum over $k$ in Eq. (\ref{eq: CGF}) is dominated by $k = {\cal O}(N)$ so that expressions for $L_k(r)$ and $M_k(r)$ have the following uniform asymptotic expansions  \cite{Temme1975}
\begin{eqnarray}
M_{ xN}(r)& = & e^{-N(r^2-x-x\log\frac{r^2}{x}) + {\cal O}(1) }\;,\quad\quad x\leq r^2, \\
M_{ xN}(r)& = & 1-e^{-N(r^2-x-x\log\frac{r^2}{x})+ {\cal O}(1)}\;,\; x\geq r^2\,.
\end{eqnarray}
To simplify the notations, we set
\be
\eta(r,x)=r^2-x-x\log\frac{r^2}{x}\;,
\ee
which is a positive function.
Inserting these asymptotic forms into Eq. \eqref{eq: CGF} and replacing the discrete sum over $k$ by an integral over $x=k/N$ we obtain the scaling form
\be
\chi_r(\mu,N)\approx N^2\Xi\left(\frac{\mu}{N},r\right)\;,
\ee
where the function $\Xi(\lambda,r)$ reads
\begin{align}\label{xi}
\Xi(\lambda,r)&=\int_0^{r^2}dx \left[\frac{1}{N}\ln\left(e^{-N\eta(r,x)}+e^{-N\lambda}\right)+\lambda r^2\right]\nn\\
&+\int_{r^2}^1 dx\left[\frac{1}{N}\ln\left(1+e^{-N(\lambda+\eta(r,x))}\right)\right]\,.
\end{align}
Since $\eta(r,x)>0$, the second integral vanishes in the large $N$ limit for $\lambda>0$. The contribution of the first integral can be expressed in terms of the solution $g(\lambda,r)$ in $x$ and in the interval $0\leq x\leq r^2$  of the equation
\be\label{Eq_eta}
\eta(r,x)=r^2-x-x\log\frac{r^2}{x}=|\lambda|\;.
\ee 
In the first integral of Eq. \eqref{xi}, the term in $\eta(r,x)$ will remain exponentially smaller than the term in $\lambda$ for $0<x<g(\lambda,r)$ while it is exponentially larger for $g(\lambda,r)<x<r^2$. This yields
\be
\Xi(\lambda,r)=\int_{g(\lambda,r)}^{r^2}dx \left[\lambda-\eta(r,x)\right]\;,\;\;\lambda\geq 0\;.
\ee
The PDF can be obtained from this scaling form of the cumulant generating function by Laplace inversion
\be
{\cal P}_r(\kappa,N)\approx \int_{\cal C}\frac{d\lambda}{2i\pi}e^{N^2\left[\lambda(\kappa-r^2)+\Xi\left(\lambda,r\right)\right]}\;,
\ee
where ${\cal C}$ is the Bromwich contour. Evaluating this integral by a saddle-point approximation finally yields the scaling form in the third line of Eq. \eqref{main_results}, where the rate function $\Psi_r(\kappa)$ is obtained for atypical fluctuations to the left of the typical regime $0\leq \kappa\leq r^2$ as
\be\label{min_lambda}
\Psi_r(\kappa)=-\min_{\lambda}\left\{\lambda(\kappa-r^2)+\Xi\left(\lambda,r\right)\right\}\;.
\ee
In particular, the saddle point $\lambda^*$ for $0\leq \kappa\leq r^2$ is such that $g(\lambda^*,r)=\kappa$. Inserting this result in Eq. \eqref{min_lambda}, we obtain 
\beeq{
\Psi_r(\kappa)&=\int_{\kappa}^{r^2} dx\left(r^2-x-x\log\frac{r^2}{x}\right)\\
&=\frac{1}{4}\left[r^4-4r^2\kappa +3\kappa^2+2\kappa^2\log\left(\frac{r^2}{\kappa}\right)\right]\,.
}
To obtain the behaviour of $\Psi_r(\kappa)$ for $r^2\geq \kappa\geq 1$, one needs to evaluate the saddle point for $\lambda^{*}\leq 0$ and it therefore requires to obtain $\Xi\left(\lambda,r\right)$ for $\lambda\leq 0$. Following a similar analysis to the case $\lambda\geq 0$, we obtain that
\be
\Xi(\lambda,r)=-\int_{r^2}^{h(r,\lambda)}dx \left[\lambda+\eta(r,x)\right]\;,\;\;\lambda\leq 0\;,
\ee
where $h(r,\lambda)$ is the solution in $x$ and for $r^2\leq x\leq 1$ of the equation \eqref{Eq_eta}. The saddle point $\lambda^*$ in this case is such that $h(r,\lambda^*)=\kappa$ which leads to the expression
\beeq{
\Psi_r(\kappa)=\frac{1}{4}\left[-r^4+4r^2\kappa -3\kappa^2-2\kappa^2\log\left(\frac{r^2}{\kappa}\right)\right]\,,
}
valid for $r^{2}\leq \kappa\leq 1$. Finally, for all values of $0\leq \kappa\leq 1$, the rate function $\Psi_r(\kappa)$ can be expressed as in Eq. \eqref{explicit_psi}, thus recovering the result of \cite{allez2014index}. Note that in the limit $\kappa=1$, the rate function $\Psi_r(\kappa=1)$ is the large deviation function of $r_{\max}$ already obtained in \cite{cunden2016large} with a similar technique, viz.
\begin{align}
\Psi_r(\kappa=1)&=-\frac{1}{N^2}\ln {\rm Prob}\left[r_{\max}\leq r\right]\\
&=-\frac{1}{4}(r^4-4r^2 +3)-\log\left(r\right)\;.
\end{align}
On the other hand, for $\kappa=0$ we obtain the large deviation function of $r_{\min}$
\be
\Psi_r(\kappa=0)=-\frac{1}{N^2}\ln {\rm Prob}\left[r_{\min}\geq r\right]=\frac{r^4}{4}\;.
\ee

%%%%%%%%%%%%%%%%%%%%%%%%%%
\section{Connection with current fluctuations in a diffusive system}
\label{sect:DerGer}
%%%%%%%%%%%%%%%%%%%%%%%%%%
Interestingly, it turns out that the scaling function describing the cumulant generating function in Eq. \eqref{chi_mu_inter2} also appears in a seemingly unrelated problem of diffusing particles on a line, studied in \cite{derrida2009current}, that we briefly describe here. Consider a set independent Brownian motions on a line, starting at $t=0$ from initial positions $x_0$ which are drawn from the following  step-like density 
\be\label{rho_x0}
\rho^{\rm BM}(x_0)=\rho_a\Theta(-x_0)+\rho_b\Theta(x_0)\;.
\ee
The system then evolves diffusively for later times $t>0$, with a diffusion constant $D$ which we set to one, for simplicity. The probability to find  a Brownian particle at a position $x$ at time $t$, having started at position $x_0$ at time $t=0$, is given by the standard propagator 
\be
P_t(x|x_0)=\frac{e^{-\frac{(x-x_0)^2}{4t}}}{\sqrt{4\pi t}}\;.
\ee
Notice that this problem has two sources of randomness: one due to the natural stochastic nature of the Brownian path, and a second one due to the initial conditions. Let us then focus on the current $Q_t$ through the origin at $x=0$, which corresponds the number of particles that have crossed the origin, up to time~$t$. If we consider the initial conditions as quenched disorder, as in the theory of disordered systems, then the natural definition of the cumulant generating function of the current ought to be (in the same spirit as Eq. (\ref{chi_mu_univ}))  \cite{derrida2009current}
\be\label{def_quenched}
\chi^{\rm q}(\mu,t) = \overline{\ln \langle e^{-\mu Q_t} \rangle_{\rm th}} \;,
\ee
where the superscript `${\rm q}$' refers to a quenched average, and  $\langle \cdots \rangle_{\rm th}$ and $\overline{(\cdots)}$ correspond to the average over the  different Brownian trajectories for fixed initial conditions, and the average over initial conditions, respectively.
It turns out that the quenched disorder is dominated by {\it typical} initial conditions, and the authors in  Ref. \cite{derrida2009current} were able to derive an exact expression for $\chi^{\rm q}(\mu,t)$ for arbitrary $\rho_a$ and $\rho_b$ [see Eq.~(\ref{rho_x0})]:
\begin{eqnarray}
\frac{\chi^{\rm q}(\mu,t)}{2\sqrt{t}}   &=& \rho_a\int_{-\infty}^0 \ln\left[\frac{1}{2}\erfc(s)+\frac{e^{-\mu}}{2}\erfc(-s)\right]ds\nn\\
&+&\rho_b\int_{0}^{\infty} \ln\left[\frac{1}{2}\erfc(-s)+\frac{e^{\mu}}{2}\erfc(s)\right]ds\;.
\end{eqnarray}
Setting $\rho_a=\rho_b=\rho$, changing variable $s\to -s$ in the first integral and gathering all the terms in the logarithm we see that $\overline{\ln \bracket{e^{-\mu Q_t}}}_{\text{th}}=2\sqrt{t}\rho\;\chi(\mu)$. Thus, in this mapping, the time $t$ plays the role of $N$, while $\rho =r$ \footnote{There is an unfortunate  misprint in Eqs. (52) and (54) in Ref.~\cite{derrida2009current}, where $t$ should be replaced by $\sqrt{t}$.}. Since the fluctuations of the integrated current in single file systems are closely related to the distribution of the displacement of a tagged particle \cite{sadhu2015large}, the CGF $\chi(\mu)$ also appears in the computation of the distribution of the displacement $X_t$ made of a tagged particle after time $t$ \cite{Krapivsky2015, cividini2017tagged}. This formal similarity between the cumulant generating function of diffusive current in the one-dimensional model and that for the number of particles in a circular disc in the two dimensional Ginibre ensemble is rather intriguing. It would be interesting to explore if there is a deeper one-to-one correspondence between these~two~problems.

%%%%%%%%%%%%%%%%%%%%%%%%%%
\section{Comparison with Monte Carlo Simulations}
\label{sect:MC}
\begin{figure}[t!]
\begin{center}
\includegraphics[width = \linewidth]{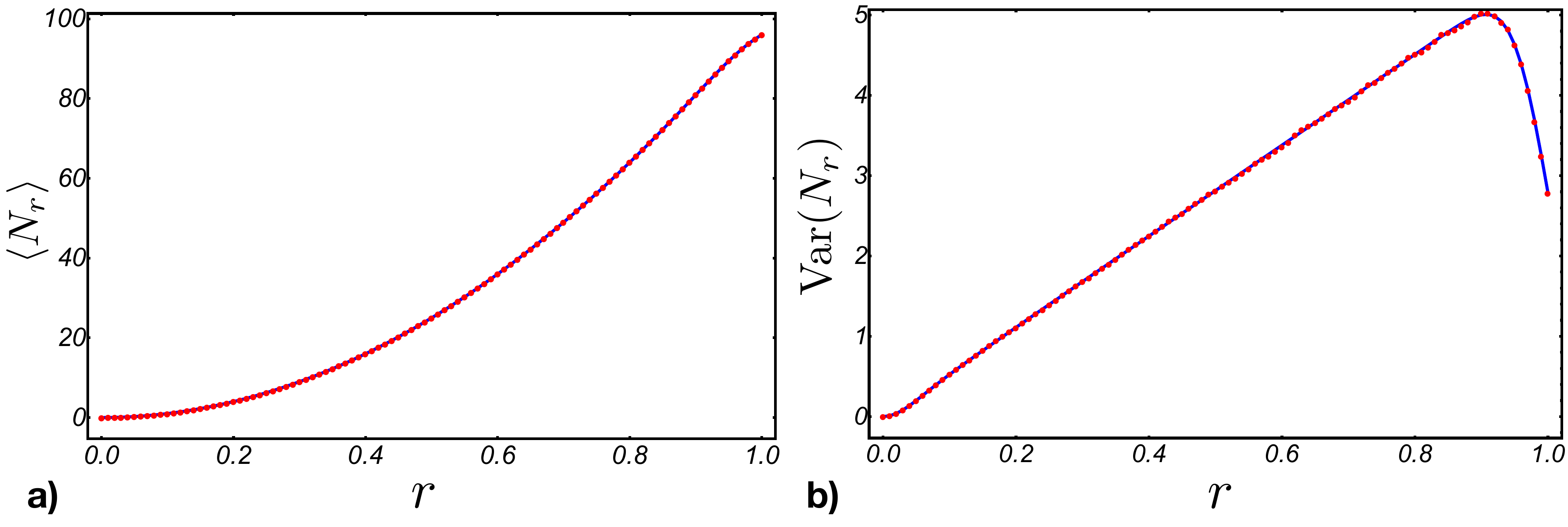}
\caption{Plot comparing the exact expressions for the first cumulants  $\bracket{N_r}$ (left panel) and  $\text{Var}(N_r)$ (right panel) of $N_r$ as a function of $r$ (solid blue lines) with Monte Carlo simulations. The latter were performed for $N=100$ and averaging over $10^5$ samples.}
\label{fig:2}
\end{center}
\end{figure}

%%%%%%%%%%%%%%%%%%%%%%%%%%
We first compare our theoretical predictions, exact for any $N$, for the mean value and variance of $N_r$, given by Eqs. (\ref{av_Nr}) and (\ref{eq:var:exact}) respectively, with Monte Carlo simulations. These observables are dominated by typical fluctuations of $N_r$ and they can be precisely computed by using a standard Metropolis algorithm to simulate the joint PDF of the Ginibre ensemble (\ref{P_joint}) with $V(z) = |z|^2$. The comparison is shown in Fig. \ref{fig:2}, for a system of size $N=100$, and, as it can be appreciated, the agreement is very good.   

To go beyond the first two moments and access the intermediate and large deviation regimes, we need to use ``importance sampling'' techniques. Here  we follow the procedure introduced in \cite{Hartmann2018}. 
The idea to generate atypical values of $N_r$ is to force  them to become typical (this is sometimes called ``reweighting''). This is achieved by introducing the following auxiliary joint PDF for the Ginibre ensemble
\beeq{
P^{(\beta)}_{\text{joint}}(z_1,\ldots,z_N)&=\frac{e^{-\beta\sum_{i=1}^N\Theta(r-r_i)}}{Z_\beta}\\
&\times\prod_{i<j}|z_i-z_j|^2\prod_{k=1}^N e^{-NV(z_k)}\,, \label{Pbias}
}
where the parameter $\beta$ plays a role similar to the inverse temperature, but in this case it can take both positive and negative values. In Eq. (\ref{Pbias}) $Z_\beta$ is a normalisation constant, i.e. the partition function of the biased Coulomb gas (which, as we will see, does not need to be computed numerically). With respect to this auxiliary joint PDF one can also derive the corresponding PDF of $N_r$, which we denote here $\mathcal{P}_r^{(\beta)}(\kappa,N)$. Clearly the original distribution $\mathcal{P}_r(\kappa,N)$ and the new one are related by the formula
\beeq{
\mathcal{P}_r(\kappa,N)=e^{\beta N\kappa}Z_\beta \mathcal{P}_r^{(\beta)}(\kappa,N)\,.
\label{eq:7}
}
The role of $\beta$ is thus to force  $\kappa$ to fluctuate around a typical value with respect to the auxiliary distribution $P^{(\beta)}_{\text{joint}}(z_1,\ldots,z_N)$, now atypical with respect to $P_{\text{joint}}(z_1,\ldots,z_N)$. We then proceed as follows: one performs a Monte Carlo simulation with respect to  $P^{(\beta)}_{\text{joint}}(z_1,\ldots,z_N)$ and constructs the histogram of $\mathcal{P}_r^{(\beta)}(\kappa,N)$.  
Using this distribution for some $\beta \neq 0$ in Eq.~\eqref{eq:7}  one gets the distribution (up to an unknown normalization constant $Z_\beta$) in the unbiased problem (i.e. for $\beta=0$) at a region of $\kappa$ which is atypical in the original problem. Thus by changing $\beta$, one can access different atypical regions of $\kappa$ values and gluing the histograms (after proper reweighting back \cite{Hartmann2018}) together,  one  obtains the distribution of $\mathcal{P}_r(\kappa,N)$. 
%This set of  histograms are glued together to obtain the distribution of $\mathcal{P}_r(\kappa,N)$ according to the formula \eqref{eq:7}. 
When varying $\beta$ it is important that the sequence of histograms do overlap, so that they can be glued together, making the whole distribution of $\kappa$ properly normalised.

\begin{figure}[t!]
\begin{center}
\includegraphics[width=1.0\linewidth]{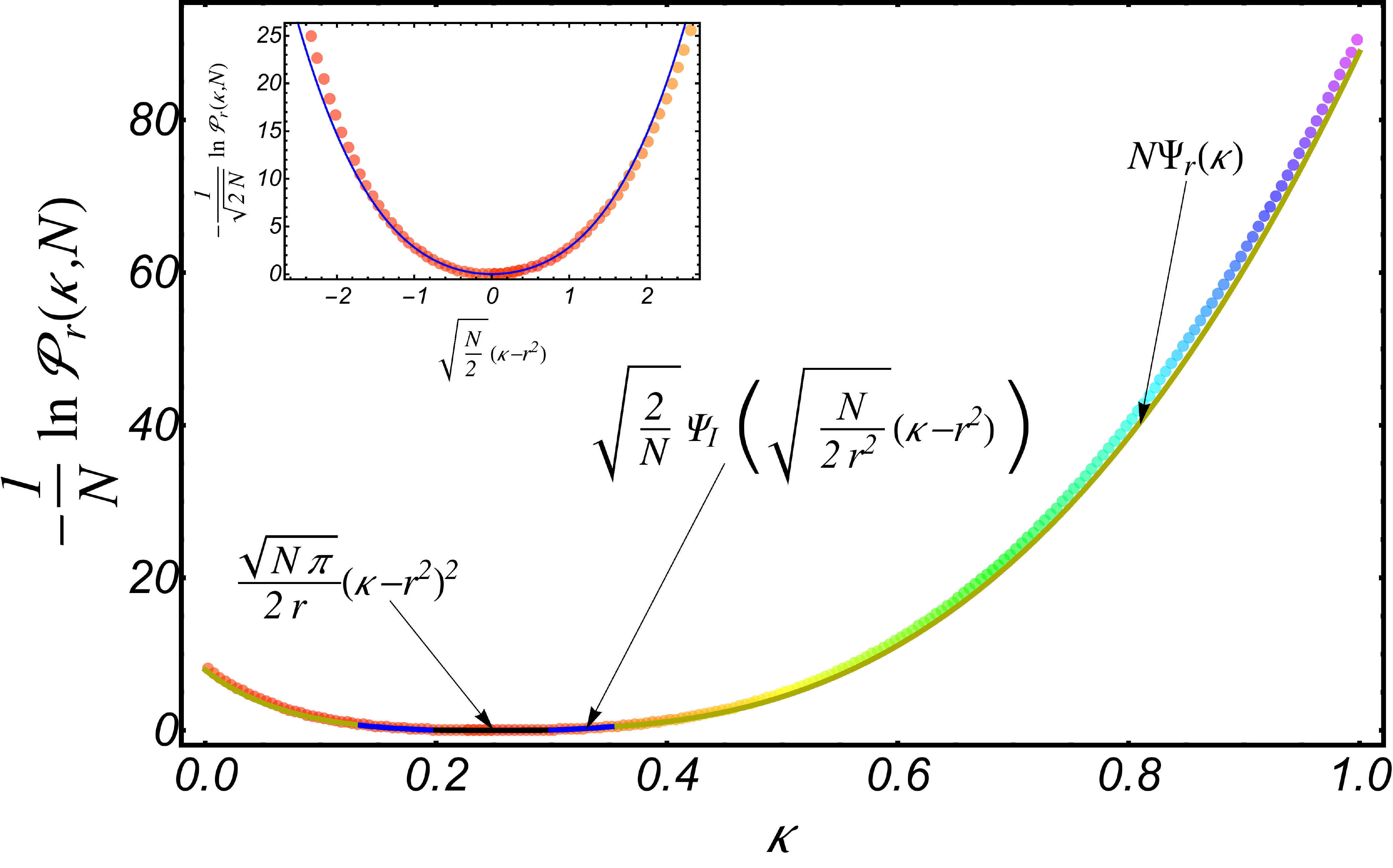}
\caption{Plot  of $-\ln\mathcal{P}_r(\kappa,N)/N$ against $\kappa$ corresponding  MC simulations (circular color markers) and the exact analytical expression of  the typical regime (solid black curve), intermediate regime (solid blue curve) and atypical regime (solid yellow curve). For the MC simulations we have used $N=500$ and values of $\beta$ ranging from $-333$  up to  $117$ in steps of $4.5$. For each temperature, Metropolis algorithm is used for $5\times 10^4$ MC steps to achieve equilibration, after which we generate samples of  $N_r$ for another $5\times 10^4$ MC steps.  The inset figure shows a more detailed plot corresponding to the intermediate regime.}
\label{fig:1}
\end{center}
\end{figure}

We have applied this reweighting Monte Carlo method to estimate $\mathcal{P}_r(\kappa,N)$ for a system of $N=500$ eigenvalues and for $w=1/2$. The results are plotted in  Fig. \ref{fig:1}, where we show the theoretical result of $-\ln \mathcal{P}_r(\kappa,N)/N$ of the three different regimes (solid black line for the typical regime, solid blue curve for the intermediate regime, and solid olive green curve for the atypical regime) together with the Monte Carlo estimates (solid circles in gradient colors representing the varying values of the inverse temperature $\beta$). Moreover, the inset of Fig. \ref{fig:1} shows a zoom 
on the intermediate regime.  As it can be appreciated, there is a fairly excellent agreement between theory and and Monte Carlo simulations.

%%%%%%%%%%%%%%%%%%%%%%%%%%
\section{Conclusions and outlook}
\label{sect:conclusions}
%%%%%%%%%%%%%%%%%%%%%%%%%%
In this paper we have  derived exact  expressions, valid for any $N$, of the distribution and all the cumulants of the number $N_r$ of  eigenvalues of  a $N \times N$ complex Ginibre matrices inside a disk of radius $r$. In the limit of large $N$, and for $0<r<1$, we showed that there are actually three different regimes of fluctuations: in addition to the typical (Gaussian) and the large deviation regimes, we have unveiled the existence of an intermediate (non-Gaussian) regime that smoothly interpolates between the two. Our main results are thus twofold. First of all, we solved a puzzling problem of matching raised by previous results obtained in Refs.  \cite{shirai2006large,allez2014index}. Secondly, our findings are  at variance with the behaviour observed in invariant models, as for instance the $\beta$-Gaussian, $\beta$-Wishart or $\beta$-Cauchy ensembles, in which there is no intermediate regime and, instead, the typical (Gaussian) and the large deviation regimes match smoothly~\cite{marino2014phase,marino2016number}. {We hope that our exact results for the FCS in the complex Ginibre ensemble will be interesting for the rather wide community working on the FCS in matrix models and related Coulomb gas systems, and more generally on random or disordered hyperuniform systems \cite{ghosh2017fluctuations, torquato2018hyperuniform}}. Besides, the results found in the present work could be experimentally relevant, as it has recently been shown that the positions of the eigenvalues of the complex Ginibre ensemble (\ref{P_joint}) are in one-to-one correspondence with the positions of fermions in a $2d$-rotating harmonic trap \cite{lacroix2019rotating}, a system which, in principle, can be realized experimentally (see e.g. \cite{schweikhard2004rapidly}).

The intermediate regime we have uncovered in our analysis  is reminiscent of the one observed for the eigenvalue with largest modulus $r_{\max}$ in the complex Ginibre ensemble \cite{cunden2016large,lacroix2018extremes}, where the existence of an intermediate regime was also found. In fact, a similar intermediate regime for $r_{\max}$ was also found for $2d$-trapped fermions, which constitutes another instance of $2d$-determinantal point process with spherical symmetry \cite{dean2017statistics,lacroix2017statistics}. Note that such a regime does not exist for the invariant ensemble where the large deviations of the largest eigenvalue matches with the typical regime of fluctuations is described by the Tracy-Widom distribution, the two being separated by a third-order phase transition (for a brief review see \cite{majumdar2014top}).

One may wonder whether such an intermediate regime for the number of eigenvalues (or particles) in a given domain also exists for other
matrix models or Coulomb gases. In particular, a problem of matching between the typical and the large deviation regimes was   recently found in \cite{rojas2018universal} for the positions of the bulk particles in the $1d$ Coulomb gas in an external convex potential. As it was pointed out in \cite{rojas2018universal},  this mismatch naturally indicates the presence of a intermediate regime, but the authors in this work were unable to mathematically describe  such regime. Oddly enough, it seems that in this particular model, there is no intermediate phase for the fluctuations of the position $r_{\max}$ of the rightmost particle \cite{dhar2017exact}. Thus it would be worth to revisit this problem and try to characterize its intermediate regime completely for bulk particles. Finally, it is natural to ask whether an intermediate regime for $N_r$ may also be found  in the real or symplectic Ginibre ensembles.

\begin{acknowledgments}
I. P. C. acknowledges support  by the programs UNAM-DGAPA-PAPIIT IA101815 and IA103417. A. K., S. N. M. and G. S. would like to acknowledge the support from the Indo-French Centre for the promotion of advanced research (IFCPAR) under Project No. 5604-2. {S. N. M acknowledges the support from the Science and Engineering Research Board (SERB, government of India), under the VAJRA faculty scheme (Ref. VJR/2017/000110) during a visit to Raman Research Institute, where part of this work was carried out}. This work was partially supported by ANR grant ANR-17-CE30-0027-01 RaMaTraF.
\end{acknowledgments}

\begin{appendix}

%%%%%%%%%%%%%%%%%%%%%%%%%%
\section{Finite $N$ results for the cumulant generating function}\label{app_CGF}
%%%%%%%%%%%%%%%%%%%%%%%%%%
We first provide a short derivation of the formula given in Eq. (\ref{P_rad}). For this purpose, we use the Vandermonde identity to obtain
\begin{align}
&\prod_{i<j}|z_i-z_j|^2=\det_{1\leq i,j\leq N} z_i^{j-1} \det_{1\leq k,l\leq N} \bar{z_l}^{k-1} \nonumber \\
&=\sum_{\sigma_1,\sigma_2 \in S_N}\sign(\sigma_1)\sign(\sigma_2)\prod_{l=1}^N z_l^{\sigma_1(l)-1}  \bar{z_l}^{\sigma_2(l)-1}\;, \label{vdm}
\end{align} 
where we used the Leibniz formula $\det_{1 \leq i,j \leq N } a_{ij}=\sum_{\sigma\in S_N}\sign(\sigma)\prod_{l=1}^N a_{l,\sigma(l)}$. Injecting this formula (\ref{vdm}) in Eq. \eqref{P_joint} and integrating over the phases $\theta_i=\arg{z_i}$, we obtain
\begin{eqnarray}
P_{\rm rad}(r_1,\cdots,r_N)&=&\int_0^{2\pi}\cdots\int_0^{2\pi} \prod_{i=1}^N r_i \, d\theta_i \, P_{\text{joint}}(z_1,\ldots,z_N) \nn \\
&=&\frac{1}{N!}\sum_{\sigma\in S_N}\prod_{k=1}^N \frac{r_k^{2\sigma(k)-1}}{h_{\sigma(k)}}e^{-Nv(r_k)}\;,\label{P_rad_app}
\end{eqnarray}
where we used the identity $\int_0^{2\pi} e^{i(m-n)\theta}d\theta=2\pi\delta_{m,n}$ and where $h_k$ is given in Eq. (\ref{h_k}). This yields the Eq. (\ref{P_rad}) given in the text.

We now compute the moment generating function which reads, by definition
\begin{eqnarray}
&&\langle e^{-\mu N_r}\rangle\\
&=&\int_{0}^{\infty}\cdots\int_{0}^{\infty} \prod_{i=1}^N dr_i P_{\rm rad}(r_1,\cdots,r_N)e^{-\mu\sum_{i=1}^N \theta(r-r_i)} \nn \;.
\end{eqnarray}
Inserting the expression for the joint PDF of the radii given in Eq. (\ref{P_rad_app}) we obtain
\begin{eqnarray}
\bracket{e^{-\mu N_r}} &=& \frac{1}{N!}\sum_{\sigma\in {\cal S}_N}\prod_{k=1}^N\int_0^{\infty}du \frac{u^{2\sigma(k)-1}}{h_{\sigma(k)}}e^{-Nv(u)-\mu \theta(r-u)}  \nn \\
&=&\prod_{k=1}^N\left[ e^{-\mu} L_k(r)+M_k(r)\right]\;,
\end{eqnarray}
where $L_k(r)$ is given in Eq. (\ref{av_Nr}) and $M_k(r) = 1 - L_k(r)$. Taking the logarithm, and using the expression for $\langle N_r \rangle$ in (\ref{av_Nr}) we obtain the result given in Eq. \eqref{eq: CGF}.

\section{Matching between the three different regimes}
\label{AppE}
%%%%%%%%%%%%%%%%%%%%%%%%%%
Let us finally check that, as stated in the text, the rate function characterising the intermediate phase interpolates between the typical and the atpypical regimes. One naturally expects to recover the typical regime for small values of $\mu$, while the large argument regime is determined by the large $|\mu|$ behavior of the function $ \chi (\mu)$. Now, in the first case, expanding $ \chi(\mu)$ for small $\mu$ we obtain:
\beeq{
 \chi (\mu)=\frac{\mu^2}{4}\sqrt{\frac{2}{\pi}}+\mathcal{O}(\mu^4)=\frac{\mu^2}{2\sqrt{2}\pi}+\mathcal{O}(\mu^4)\,,
}
where we have used that
\beeq{
\int_0^\infty  du \, \erfc\left(u\right)\erfc\left(-u\right)=\sqrt{\frac{2}{\pi}}\,.
}
This implies that the rate function $\Psi_I(x)$ controlling the intermediate phase  becomes
\begin{eqnarray}
\Psi_I(x)&=&-\min_{\mu}\left\{\mu x+\chi(\mu)\right\} \approx-\min_{\mu}\left\{\mu x+\frac{\mu^2}{2\sqrt{2}\pi}\right\} \nn \\
&\approx&\sqrt{\frac{\pi}{2}}x^2 \;, \; {\rm as} \;\;  x  \to 0 \;,
\end{eqnarray}
where we have used that the minimum is achieved at $\mu^*=-x\sqrt{2\pi}$. Thus we have that $P_w^{(N)}(y)\approx e^{-\sqrt{2N}r \sqrt{\frac{\pi}{2}}x^2}$ and recalling that $x=\sqrt{N/(2r^2)}(\kappa-r^2)$ we finally obtain
\beeq{
{\cal P}_r(\kappa,N)\approx e^{-N^{\frac{3}{2}}\frac{\sqrt{\pi}}{2r}(\kappa-r^2)^2}
}
 which corresponds to the typical regime of fluctuations, as given in the first line of Eq. \eqref{main_results}. 
 
On the other hand, for large values of $|\mu|$, we first realise that the integrand appearing in $\chi(\mu)$ in Eq. (\ref{chi_mu_inter2}) behaves differently whenever
 $s<\sqrt{|\mu|}$ or $s>\sqrt{|\mu|}$. In turns out that only the first region $s < \sqrt{|\mu|}$ contributes, at leading order for large $|\mu|$, yielding
\beeq{
 \chi(\mu)&\approx  \int_0^{\sqrt{|\mu|}} ds \log\left[1+\frac{2}{\sqrt{\pi}s} e^{|\mu|-s^2}\right]\\
 &\approx  \int_0^{\sqrt{|\mu|}} du \log\left[\frac{2}{\sqrt{\pi}u} e^{|\mu|-u^2}\right]\approx \frac{2}{3}|\mu|^{3/2}\\
%  &=\sqrt{|\mu|} \log\left[\frac{2}{\sqrt{\pi}}\right] \approx \frac{2}{3}|\mu|^{3/2}\\
%  &+\left(\sqrt{\mu}+\frac{2}{3}\mu^{3/2}+\sqrt{\mu}\log(\sqrt{\mu})\right)\\
% &\sim \frac{2}{3}\mu^{3/2}\,.
}
Thus the expression for $\Psi_I(x)$ is obtained by solving the following minimization problem:
\beeq{
\Psi_I(x)&=-\min_{\mu}\left\{\mu x+\chi(\mu)\right\}\\
&\approx-\min_{\mu}\left\{\mu x+\frac{2}{3}|\mu|^{3/2}\right\}\\
&\approx\frac{|x|^3}{3}\,, \, {\rm as} \; x \to \pm \infty \;,
}
where in the last step we have used that the value of $\mu$ which minimizes the above expression is $\mu^* = - {\rm sign}(x)x^2$ (which is positive for negative $x$ and negative for positive $x$). Finally, in terms of the original variables $x=\sqrt{N/(2r^2)}(\kappa-r^2)$, the tail of the intermediate regime reads
\beeq{
{\cal P}_r(\kappa,N)&\approx \exp\left(-\sqrt{2N}r\frac{|x|^3}{3}\right)\\
&\approx\exp\left(-N^2\frac{|\kappa-r^2|^3}{6 r^2}\right) \;,
}
which exactly coincides with the small argument behavior of the large deviation regime (i.e. the third line in (\ref{main_results})) since the large deviation function exhibits exactly the same cubic behaviour (\ref{cubic_behavior}). We thus conclude that the intermediate regime matches perfectly the typical and atypical regimes.

\end{appendix}

\newpage

\bibliography{biblio}

\end{document}